\documentclass[lettersize,journal]{IEEEtran}

\usepackage{amsmath,amsfonts}
\usepackage{algorithm}
\usepackage{algpseudocode}
\usepackage{array}
\usepackage{textcomp}
\usepackage{stfloats}
\usepackage{url}
\usepackage{verbatim}
\usepackage{graphicx}
\usepackage{cite}
\usepackage{xcolor}
\usepackage{lipsum}
\usepackage{listings}
\usepackage{tabularx}
\usepackage{array}
\usepackage{amssymb}
\usepackage[table]{xcolor}

\usepackage[skip=4pt]{caption}

\usepackage{titlesec}

\titlespacing*{\section}{0pt}{1ex}{0.5ex}
\titlespacing*{\subsection}{0pt}{1ex}{0.5ex}
\titlespacing*{\subsubsection}{0pt}{1ex}{0.5ex}

\usepackage[]{cite}
\usepackage[a4paper,top=2.54cm,bottom=2.54cm,left=2.54cm,right=2.54cm,headheight=16pt]{geometry}
\usepackage{amsmath}
\usepackage[autostyle=true]{csquotes} 
\begin{document}


\title{ Exploring LLM in Semantic Communication for V2X Networks}



\author{Sihem Bakri \IEEEauthorrefmark{1}, Navdeep Singh \IEEEauthorrefmark{1}, Nicola Marchetti \IEEEauthorrefmark{1}, 


\IEEEauthorblockA{\IEEEauthorrefmark{1} Trinity College Dublin, Dublin, Ireland;}

\IEEEauthorblockA{Emails: \IEEEauthorrefmark{1}sbakri@tcd.ie, \IEEEauthorrefmark{1}nasingh@tcd.ie, \IEEEauthorrefmark{1}nicola.marchetti@tcd.ie
}

}


\maketitle

\begin{abstract}
\thispagestyle{empty}
The rapid growth of connected and autonomous vehicles has created a demand for more efficient and intelligent communication systems. Traditional Vehicle-to-Everything (V2X) networks rely on transmitting raw sensor data, leading to high bandwidth usage and redundant information exchange. To address this, we propose a semantic communication framework that integrates a Large Language Model (LLM) with graph-based knowledge representation, to transmit only high-level, meaningful messages instead of raw data.
Within this framework, the LLM performs semantic transformation, converting structured sensor inputs 
 into concise natural language messages that describe context and intent. \textcolor{black}{It also generates high-level control decisions based on shared situational awareness across the V2X network.}
A multilane traffic simulation was developed to compare semantic and non-semantic modes in terms of bandwidth usage. 
\textcolor{black}{Results show an average 33.54\% reduction in data transmission and illustrate context-aware coordination in representative scenarios.}
\end{abstract}




\section{Introduction} 
\label{sec:introduction}

The rapid advancement of autonomous vehicles has increased the need for advanced communication systems that can support real-time coordination, decision-making, and efficient resource utilization across vehicular networks. Vehicle-to-Everything (V2X) communication enables interaction among vehicles, infrastructure, pedestrians, and cloud services, which form the backbone of next-generation Intelligent Transportation Systems (ITS). 
However, traditional syntactic communication models, where raw sensor data is exchanged without interpretation, are becoming inefficient due to their high bandwidth usage, data redundancy, and limited contextual understanding.
Semantic communication has recently emerged as a promising alternative ~\cite{SC_overview}. Instead of transmitting unfiltered sensor data, semantic communication allows agents to exchange only the meaningful and goal-relevant content of information. For example, rather than sharing exhaustive LiDAR (Light Detection and Ranging) data streams, a vehicle could send a compact message such as “vehicle braking ahead in lane 2.” This approach reduces bandwidth consumption and improves system performance by providing interpretable and goal-oriented information instead of large, low-level data packets.


\setlength{\parskip}{0pt}
\textcolor{black}{ To support semantic communication in V2X systems, reasoning mechanisms are needed to interpret, generate, and contextualize semantic messages. Large Language Models (LLMs), such as OpenAI’s GPT-4, offer strong capabilities in natural language understanding and contextual inference, which make them promising candidates for this role. When integrated into vehicular networks, LLMs can help interpret environmental state information, infer relationships among agents, and generate high-level control decisions based on shared situational context. However, deploying such models in V2X settings raises practical design challenges, including maintaining consistent shared context across multiple vehicles and coordinating decisions in dynamic multi-agent environments. Motivated by these considerations, this paper proposes a system-level V2X architecture that combines LLM-based semantic reasoning with a centralized graph memory. In this system, each vehicle transmits semantic summaries of its local environment to a central reasoning server. The server uses an LLM to infer control logic informed by a shared knowledge graph that encodes vehicle positions, semantic relationships (e.g., following, overtaking, slowing down), and relevant traffic context.}

The main objectives of this paper are to:
\begin{itemize}
\item Design a semantic communication framework for multi-agent V2X systems.
\item \textcolor{black}{Integrate LLM-based reasoning to interpret semantic messages and support context-aware decision-making.}
\item Implement a graph-based persistent memory to maintain shared context across vehicles.
\item \textcolor{black}{  Evaluate the system in a simulation environment, focusing on communication efficiency and analyzing decision behavior through scenario-based validation.}
\end{itemize}

The rest of this paper is structured as follows. Section II provides background information on semantic communication and LLMs, discusses their roles in V2X networks, and reviews related work. Section III presents the proposed system model and architecture. Section IV presents and analyzes the experimental results. Finally, Section V concludes the paper with the key findings and suggestions for future research.

\section{Background and Related Work}

Vehicle-to-Everything (V2X) refers to the communication network between vehicles and other entities, including other vehicles (V2V), infrastructure nodes or roadside infrastructure (V2I), pedestrians (V2P), and other networks (V2N). 
By enabling seamless interaction between vehicles and their surroundings, V2X plays a critical role in shaping next-generation autonomous vehicular systems. 
Through V2X, vehicles can broadcast safety-critical data such as speed, position, acceleration, and braking status. This exchange enhances cooperative maneuvers like platooning, lane merging, and collision avoidance. With the adoption of standardized protocols such as LTE-PC5
and DSRC \cite{DSRC_ltepc5_comp},
V2X has evolved to support decentralized message dissemination in high-mobility environments, while emerging 6G paradigms aim to integrate semantic awareness and ultra-low latency for predictive and goal-aligned driving strategies.

Several studies have explored the concept of V2X communication; for instance, \cite{rahim2023v2x, v2x} present comprehensive overviews of enabling technologies for 6G-V2X systems. The authors highlight that 6G will play a pivotal role in enhancing the responsiveness and context awareness of V2X communications through advanced physical layer techniques, intelligent signal processing, and integration with AI-driven control systems. Machine learning algorithms, in particular, are expected to handle the dynamic and heterogeneous nature of vehicular networks by optimizing communication resources and enabling adaptive coordination among agents. \textcolor{black}{ 
However, these schemes largely operate at the syntactic level of information, which can be suboptimal under stringent latency, reliability, and bandwidth constraints.
Motivated by these limitations, recent works have begun to investigate semantic-aware approaches for V2X.} 

\subsection{Semantic Communication}

Semantic communication departs from conventional syntactic transmission by focusing on the meaning and purpose of the message rather than the content of the raw data. In this context, Liu et al.~\cite{liu2024survey} characterize semantic communication systems as comprising three primary components: semantic encoders, knowledge bases, and semantic decoders. 
Getu et al.~\cite{getu2023scfuture} emphasize the role of semantic communication as a key enabler in 6G systems. In addition, the surveys in \cite{liu2024survey} and \cite{getu2023scfuture} identify several open challenges in semantic communication including the lack of standardized semantic accuracy metrics and difficulties in modeling shared knowledge across heterogeneous agents. These concerns are directly relevant to multi-agent vehicular systems, where dynamic context and distributed intelligence complicate semantic agreement.
From a theoretical standpoint, Xin et al.~\cite{xin2024semantic} provide an analysis of semantic entropy, rate-distortion functions, and semantic channel capacity. Their work bridges the gap between Shannon's classical information theory and the requirements of meaning-aware communication systems. They advocate for the development of coding schemes that preserve semantic integrity while minimizing resource usage.
In terms of the security aspects of semantic networks, Guo et al.~\cite{guo2024semcomnet} propose a three-layer architecture that categorizes privacy threats and propose defense mechanisms for multi-agent semantic communication networks (SemComNet). Their analysis includes privacy-preserving reasoning, adversarial robustness, and secure message interpretation, which are crucial considerations in mission-critical V2X scenarios.
\textcolor{black}{
In the V2X context, Lyu et al.~\cite{lyu2024semv2x} introduce a hierarchical semantic V2X architecture in which vehicles use semantic encoders, share a common knowledge base, and are evaluated using metrics such as Age of Information (AoI). They show that compact semantic messages (e.g., “vehicle braking ahead”, “lane blocked”) can replace raw sensor or video data, while still supporting cooperative driving, traffic awareness, and intelligent mobility.} 

\textcolor{black}{ 
While these semantic communication approaches show that focusing on task-relevant meaning can improve V2X performance, they often rely on task-specific models with limited flexibility. Recent progress in LLMs offers a more general way to represent and reason about semantics, and that is what we are introducing in our work.}

\subsection{Large Language Model}

LLMs are advanced deep learning models trained on vast amounts of textual data to perform a wide range of natural language processing tasks. These models, typically based on transformer architectures, learn statistical patterns in language to generate coherent and contextually appropriate responses ~\cite{brown2020language,openai2023gpt4}.
Unlike traditional models that require fine-tuning for each specific task, LLMs can operate in zero-shot, one-shot, or few-shot settings. This is made possible by prompt-based learning, where the model is guided by natural language instructions. GPT-4, in particular, shows improved reasoning, factual accuracy, and multilingual capabilities compared to its predecessors~\cite{openai2023gpt4}.

\vspace{0.9em}

\subsubsection{LLMs in Communication Systems}

\textcolor{black}{ LLMs are emerging as useful tools in communication systems due to their ability to process context-aware information. They can support traffic management, coordination, and real-time decision-making in communication networks.}
LLMs are a recent advancement, and only a few research works have begun to explore their applications.
In this context,  Zhou et al.~\cite{zhou2024telecom} present a detailed overview of how LLMs can be applied in telecommunications, namely: traffic classification and prediction, network resource optimization, user interaction and support, and intelligent decision-making in large-scale systems.
Ameur et al.~\cite{ameur2024xrl} focus on using LLMs to explain the behavior of deep reinforcement learning (DRL) models in 6G networks. In their framework, LLMs help translate internal decision processes into understandable language, which improves transparency and helps engineers, operators, and users better understand system behavior. 
Friha et al.~\cite{friha2024edge} explore the integration of LLMs into edge computing environments. In these setups, LLMs are used to process and analyze data closer to where it is generated such as in vehicles or roadside units. 
Ullah et al.~\cite{ullah2024smartcities} examine how LLMs can support smart cities, with use cases in smart transportation, public safety, and energy management. Their study highlights the challenges of multilingual communication, variations in language formats, and ethical considerations when using LLMs in public systems. 
Long et al.~\cite{long2024optimization} propose an intelligent 6G network architecture that uses LLMs for performance optimization. Their system uses LLMs to interpret data patterns and support automatic network health checks, traffic control, and operational planning.

\vspace{0.9em}
\subsubsection{LLMs for Semantic Communication}


\textcolor{black}{ LLMs are promising for semantic communication because they can process meaning and context efficiently. This makes them useful in systems focused on exchanging meaning rather than raw data.}
Very few research works have introduced LLMs for semantic communication. 
Jiang et al.~\cite{jiang2024lammsc} introduce a framework called LAM-MSC (Large AI Model-based Multimodal Semantic Communication), which combines LLMs with multimodal data namely visual, textual, and sensor modalities to improve communication quality. 
Kalita et al.~\cite{kalita2024iotllm} propose an LLM-powered semantic communication system for edge-based IoT networks. They present a modular framework where semantic encoding and decoding are guided by LLMs, making the transmission more intelligent and reducing the exchange of unnecessary data. 
These frameworks demonstrate how LLMs can help improve semantic representation and context modeling in communication systems. The ability to personalize meaning through knowledge bases and maintain semantic consistency over noisy channels makes LLMs especially useful in environments with dynamic, real-time data such as vehicle networks.


\vspace{0.9em}
\subsubsection{LLMs in V2X communication}
Recent efforts in vehicular networks have shifted towards incorporating semantic reasoning capabilities using LLMs, particularly to enhance decision-making and reduce data transmission burdens.
Zhang et al.~\cite{zhang2025embodied} propose a framework combining LLMs with DRL in an embodied AI-enhanced vehicular network. The system utilizes the LLAVA (Large Language and Vision Assistant) model to convert visual data into semantic representations and employs Proximal Policy Optimization (PPO) for real-time navigation decisions. Their results indicate significant reductions in bandwidth usage and improvements in Quality of Experience (QoE) compared to traditional DRL agents. 
Another approach in~\cite{hybrid2024reasoning} implements a hybrid reasoning system where LLMs perform natural language interpretation of driving scenarios to assist rule-based planning in complex environments. This design showcases how LLMs can enhance contextual understanding and decision support in vehicular control systems. Specifically, the authors propose a method for real-time transformation of traffic scenes into descriptive text and structured knowledge graphs (KGs).



\begin{table*}[t]
\centering
\captionsetup{labelfont={color=black},textfont={color=black}}
\color{black}
\arrayrulecolor{black}
\caption{Comparison of representative related works and our proposed framework.}
\label{tab:related_work_comparison}
\renewcommand{\arraystretch}{1.2}
\setlength{\tabcolsep}{4pt}
\footnotesize
\begin{tabularx}{\textwidth}{|p{3.3cm}|p{1.8cm}|c|c|c|p{2cm}|X|}
\hline
\textbf{Work} & \textbf{Context} & \textbf{LLM} & \textbf{KG} & \textbf{Semantic} & \textbf{Arch.} & \textbf{Key Difference} \\
\hline
DSRC vs LTE-V2X~\cite{DSRC_ltepc5_comp}
& V2X
& $\times$
& $\times$
& $\times$
& Distributed
& Focuses on communication technologies and congestion control without semantic abstraction or LLM-based reasoning. \\
\hline
6G V2X overviews~\cite{rahim2023v2x,v2x}
& V2X
& $\times$
& $\times$
& $\times$
& Distributed
& Discuss enabling technologies and challenges for future V2X systems, but do not address semantic communication or LLM-driven reasoning. \\
\hline
Semantic communication surveys~\cite{liu2024survey,getu2023scfuture,xin2024semantic,guo2024semcomnet}
& Communication
& $\times$
& $\times$
& $\checkmark$
& Mixed
& Focus on semantic information exchange and its theory, security, and challenges, without LLM-based decision-making for vehicular coordination. \\
\hline
Semantic V2X~\cite{lyu2024semv2x}
& V2X
& $\times$
& $\times$
& $\checkmark$
& Distributed
& Applies semantic communication to V2X, but does not integrate LLM-based reasoning or persistent graph memory. \\
\hline
LLMs in communication systems~\cite{zhou2024telecom,ameur2024xrl,friha2024edge,ullah2024smartcities,long2024optimization}
& Communication
& $\checkmark$
& $\times$
& $\times$
& Central/Edge
& Use LLMs for telecom intelligence, explainability, edge processing, and network optimization, but not for semantic V2X coordination. \\
\hline
LLMs for semantic communication~\cite{jiang2024lammsc,kalita2024iotllm}
& Communication
& $\checkmark$
& $\times$
& $\checkmark$
& Central/Edge
& Combine LLMs with semantic communication, but are not focused on cooperative multi-vehicle V2X coordination with shared memory. \\
\hline
Embodied AI-enhanced vehicular networks~\cite{zhang2025embodied}
& Driving/V2X
& $\checkmark$
& $\times$
& Partial
& Distributed
& Uses LLMs with DRL for vehicular decision-making and semantic scene interpretation, but does not incorporate persistent graph memory or semantic V2X message exchange. \\
\hline
Hybrid reasoning for autonomous driving~\cite{hybrid2024reasoning}
& Driving
& $\checkmark$
& $\checkmark$
& $\times$
& Distributed
& Uses LLM-based hybrid reasoning with structured driving context in CARLA for brake/throttle decision support, but does not address semantic V2X communication or shared multi-vehicle graph memory. \\
\hline
\textbf{Our Work}
& V2X
& $\checkmark$
& $\checkmark$
& $\checkmark$
& Centralized
& Integrates semantic message generation, LLM-based reasoning, and persistent graph memory for cooperative V2X coordination. \\
\hline
\end{tabularx}
\arrayrulecolor{black}
\end{table*}

In contrast to previous studies that focus on LLM-assisted perception, semantic encoding, or hybrid LLM–DRL decision-making, \textcolor{black}{this work integrates LLM-driven semantic reasoning with a centralized knowledge graph (KG) in V2X networks.} This design enables persistent shared situational awareness across multiple vehicles, supporting coordinated decision-making while reducing communication overhead.

\textcolor{black}{ Table~\ref{tab:related_work_comparison} summarizes representative works discussed above in terms of their use of semantic communication, LLM-based reasoning, and graph-based contextual modeling. As shown, prior studies generally emphasize either communication efficiency or intelligent decision-making in isolation. In contrast, our framework combines semantic message generation, LLM-based reasoning, and persistent graph memory within a unified V2X coordination architecture.}

\section{System architecture}
\label{sec:Model}

This work presents a multi-lane car simulation that models autonomous vehicle behaviour in a dynamic, multi-agent road environment. 
The system integrates GPT-based decision-making, real-time semantic communication, and graph-based data storage to create an efficient and responsive V2V simulation. This allows autonomous vehicles to perceive their surroundings, interpret their environment, and intelligently respond to obstacles and other moving vehicles.
\begin{figure}[ht]
    \centering
    \includegraphics[width=0.450\textwidth]{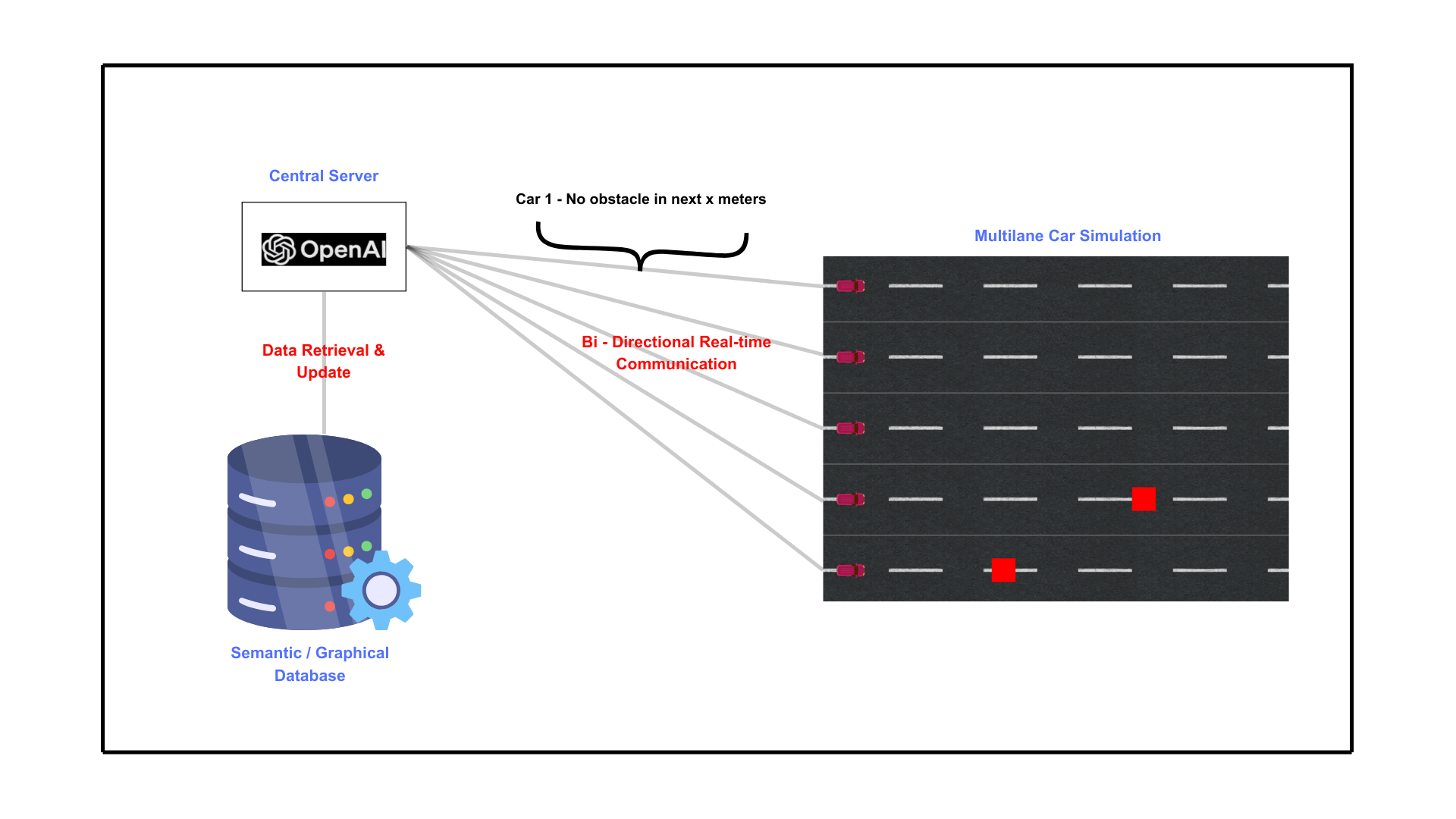}
    \caption{Overview of the System Architecture.}
    \label{fig:system_architecture}
\end{figure}


The architecture comprises the following main components as shown in Figure \ref{fig:system_architecture}: 

\subsection{LLM:} 
One of the most important components of the system is the LLM OpenAI’s GPT, which is responsible for making all intelligent decisions that guide the movement of each car in the simulation.
 For each simulation step, the server constructs a structured description of the current traffic state, including each car’s lane, distance to obstacles, neighbouring vehicles, and relevant history retrieved from the database (e.g., past collision-risk events). Based on this context and on explicit safety and performance constraints (collision avoidance, safe gaps, progress along the road), the module returns clear, high-level commands such as “MOVE FORWARD 10” (i.e that the vehicle should advance 10 units along its lane while maintaining safety limits) or “CHANGE LANE RIGHT” for the addressed vehicle. It can also propose updates to semantic relations (e.g. FOLLOWING, SLOW DOWN) that are written back to the database.

\subsection{Graph database:} 
The graph database, implemented in Neo4j \cite{neo4j}, provides long-term and short-term memory for the simulation. Neo4j is a graph database management system that stores data as nodes and relationships, enabling efficient querying and analysis of highly connected, graph-structured information. In our case, it stores lanes, vehicles, obstacles, and their relationships as nodes and edges (for example, a relationship such as VEHICLE A → FOLLOWING → VEHICLE B indicating that Vehicle A and Vehicle B are both in same lane), together with time-stamped movement commands and inferred semantic labels. As the simulation runs, each action and resulting position is recorded, producing a history of trajectories and interactions. When the server receives a new vehicle state, it queries the database for approximate positions, recent actions, neighbouring cars, and lane-change safety checks, and then passes this information to the decision module.



\subsection{System communication network:} 
\textcolor{black}{System communication relies on a TCP/IP client–server model. Each vehicle operates as a client that sends its current state (e.g., lane index, detected obstacles, and nearby vehicles) to the central server over a reliable TCP socket. The server processes all incoming messages, determines appropriate control commands for each vehicle, and sends the responses back over the same connections. Separate threads handle individual connections, allowing multiple vehicles to be updated concurrently. Vehicles do not communicate directly; all coordination is mediated by the server, which has access to the full global state and database.}

\setlength{\parskip}{0pt}

\textcolor{black}{
The centralized architecture is intentionally adopted to maintain a consistent global context and shared graph-based memory across vehicles, enabling coordinated semantic decision-making and facilitating evaluation in a controlled environment.}

\subsection{Semantic communication:} 
Semantic communication is realised by exchanging structured, meaning-oriented messages rather than raw sensor values. Instead of only sending distances or speeds, the server and database maintain higher-level relations such as whether a lane change is hazardous, whether a car is following another too closely, or whether a slowdown is required. These semantics are encoded as labels and relationships in the graph (e.g. FOLLOWING, SLOW DOWN, SPEED UP) and are included in the context provided to the decision module. This reduces ambiguity and ensures that commands are based on task-relevant information.

\subsection{Virtual V2V:} 
Virtual V2V communication is achieved through this centralized, semantics-aware design. While vehicles do not broadcast directly to each other, the server aggregates their individual reports, queries the database for nearby vehicles and recent behaviours, and then issues commands that implicitly coordinate their actions. In this way, relationships such as following distance, relative speed, and lane safety are maintained as graph structures and updated over time, yielding a history-informed and scalable approximation of V2V interaction suitable for studying coordinated autonomous driving.

\vspace{0.3cm}
\textcolor{black}{
\noindent\textbf{Real-world applicability:}
The proposed framework is a proof-of-concept coordination layer for V2X systems that can run at the network edge to enable local vehicle coordination with reduced overhead. It is compatible with LTE-V2X and future 6G systems, and complements existing perception and control modules. While currently centralized, future distributed or hierarchical versions may improve scalability and robustness.}
\setlength{\parskip}{0pt}

\vspace{0.23cm}
\textcolor{black}{
\noindent\textbf{Latency and real-time considerations:} LLM-based reasoning may introduce inference delays, depending on model size and deployment. In practice, this can be reduced through edge deployment and lightweight models. A full quantitative latency evaluation is beyond the scope of this prototype and is left for future work.
}

\section{Results }
\label{sec:Result}



In this section, we evaluate the proposed framework in terms of communication efficiency in a V2V scenario. The semantic framework is compared with a non-semantic baseline to determine whether semantic messaging reduces bandwidth while preserving coordination-relevant information.

\textcolor{black}{Experiments are conducted in a custom Python/Pygame simulator with five autonomous vehicles connected to a centralized server. Wireshark is used to measure bandwidth consumption and message rates under semantic and non-semantic communication modes.}

\vspace{0.3cm}
\textcolor{black}{
\noindent\textbf{Implementation details:} The evaluation uses five vehicles with persistent TCP connections to a central server, and each run lasts about 600 seconds. GPT-4o is used for semantic reasoning based on structured vehicle-state inputs, including lane position, obstacles, nearby vehicles, estimated positions, command history, and lane-change constraints. The model outputs control commands and semantic graph updates, while detailed latency optimization is left for future work.
}

\subsection{Bandwidth Usage Analysis}
\label{bandwidth-usage-analysis}

\textcolor{black}{ To compare semantic and non-semantic communication in our architecture, we analyze the amount of data transmitted from each vehicle to the server.} In the non-semantic setup, each vehicle transmits raw environmental data structured as LIDAR-like JSON, including null/populated range vectors and lane-position metadata (see Listing 1).
\begin{lstlisting}[caption={Sample LIDAR-Style JSON Payload}, captionpos=b]
{
  "frame_id": "car_1_lidar",
  "range_min": 0,
  "range_max": 120,
  "ranges": [
    null, null, null, null, 74.8, 74.3, 73,
    74.3, 74.8, null, null, null, null
  ],
  "lane_position": 1
}
\end{lstlisting}

In the semantic setup, vehicles instead communicated abstract, compressed/semantic messages that described their environment in natural language (see Listing 2).
\begin{lstlisting}[caption={Sample Semantic (Natural Language) Payload}, captionpos=b]
Car 3 No obstacle in next 120 meters, No car in next 120 meters, LANE 3
\end{lstlisting}

We ran each simulation mode for about 600 seconds and captured all TCP traffic with Wireshark. For both semantic and non-semantic configurations, we analyzed the packet- and byte-level traces from the five vehicles to the server and computed per-stream transmission rates (in bits per second) to quantify overall bandwidth usage.

\begin{figure}[H]
\centering
\includegraphics[width=0.45\textwidth]{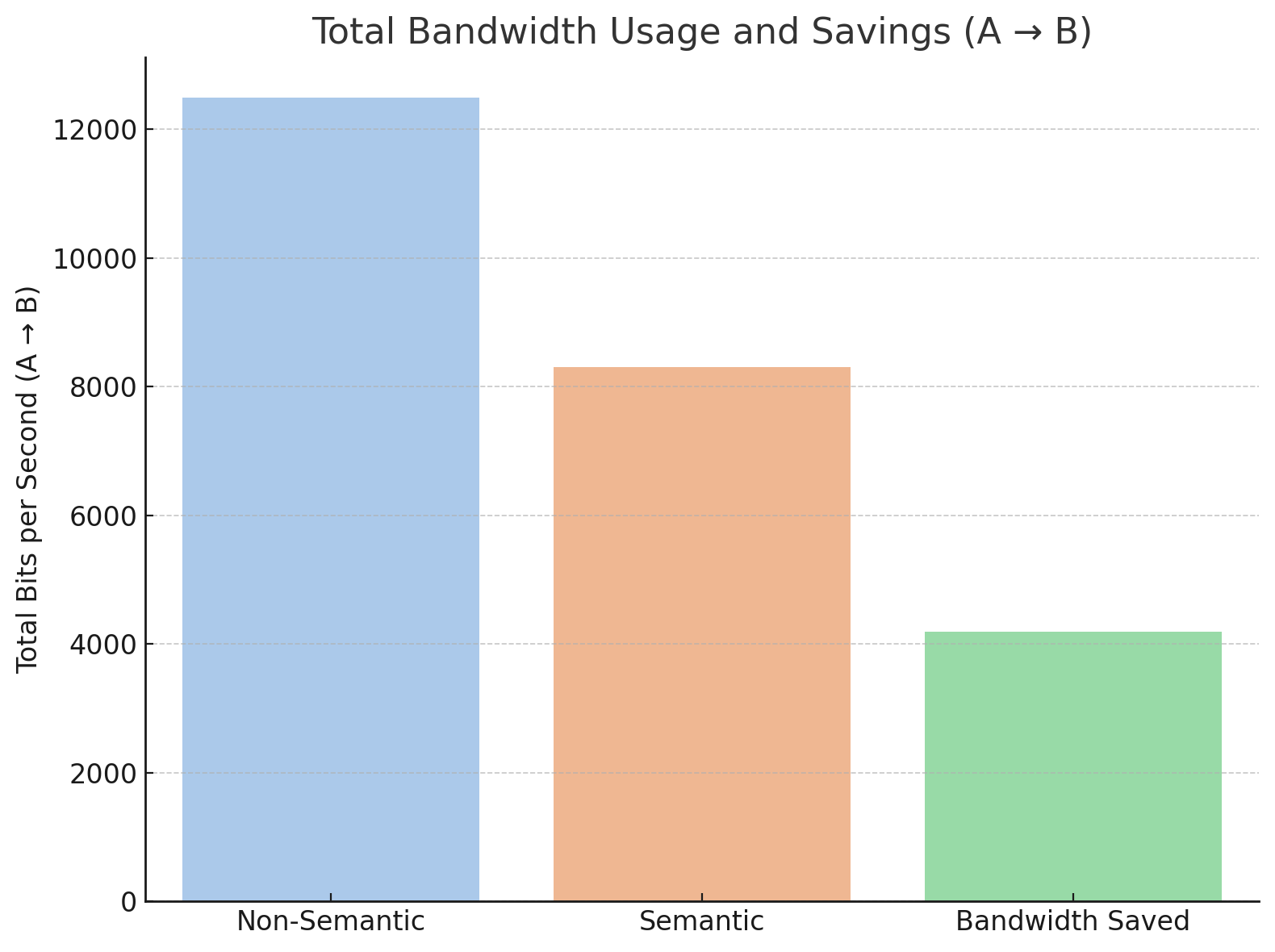}
\caption{Total Bandwidth Usage in Semantic and Non-Semantic Modes (Vehicle to Server)}
\label{fig:bandwidth-total}
\end{figure}

Figure.~\ref{fig:bandwidth-total}  shows the aggregated bandwidth usage across all TCP streams. In the non-semantic mode, the system transmitted approximately \textbf{12,498 bits per second}. In contrast, the semantic system required only \textbf{8,305 bits per second}, resulting in a total reduction of \textbf{4,193 bits per second}.
This represents a bandwidth saving of approximately \textbf{33.54\%}. The observed reduction is primarily due to the model's ability to express relevant context and intent using shorter, more meaningful messages, thereby eliminating redundant or unnecessary data transmissions.

\subsection{Stream-by-Stream Comparison}
To complement our total bandwidth analysis, we examined the data transmission rate for each individual TCP connection between a vehicle and the server. Each vehicle maintained a dedicated connection throughout the 600-second simulation, resulting in five concurrent data streams.
In the non-semantic configuration, each vehicle transmitted structured LIDAR-like JSON objects, which included 13-element `ranges` arrays along with lane position metadata. These verbose and sparsely populated messages resulted in high transmission overhead. In contrast, the semantic setup relied on short natural language statements, such as ``Car 3 – No obstacle in next 120 meters, No car in next 120 meters, LANE 3," which conveyed the same information more concisely.

\begin{figure}[H]
\centering
\includegraphics[width=0.4\textwidth]{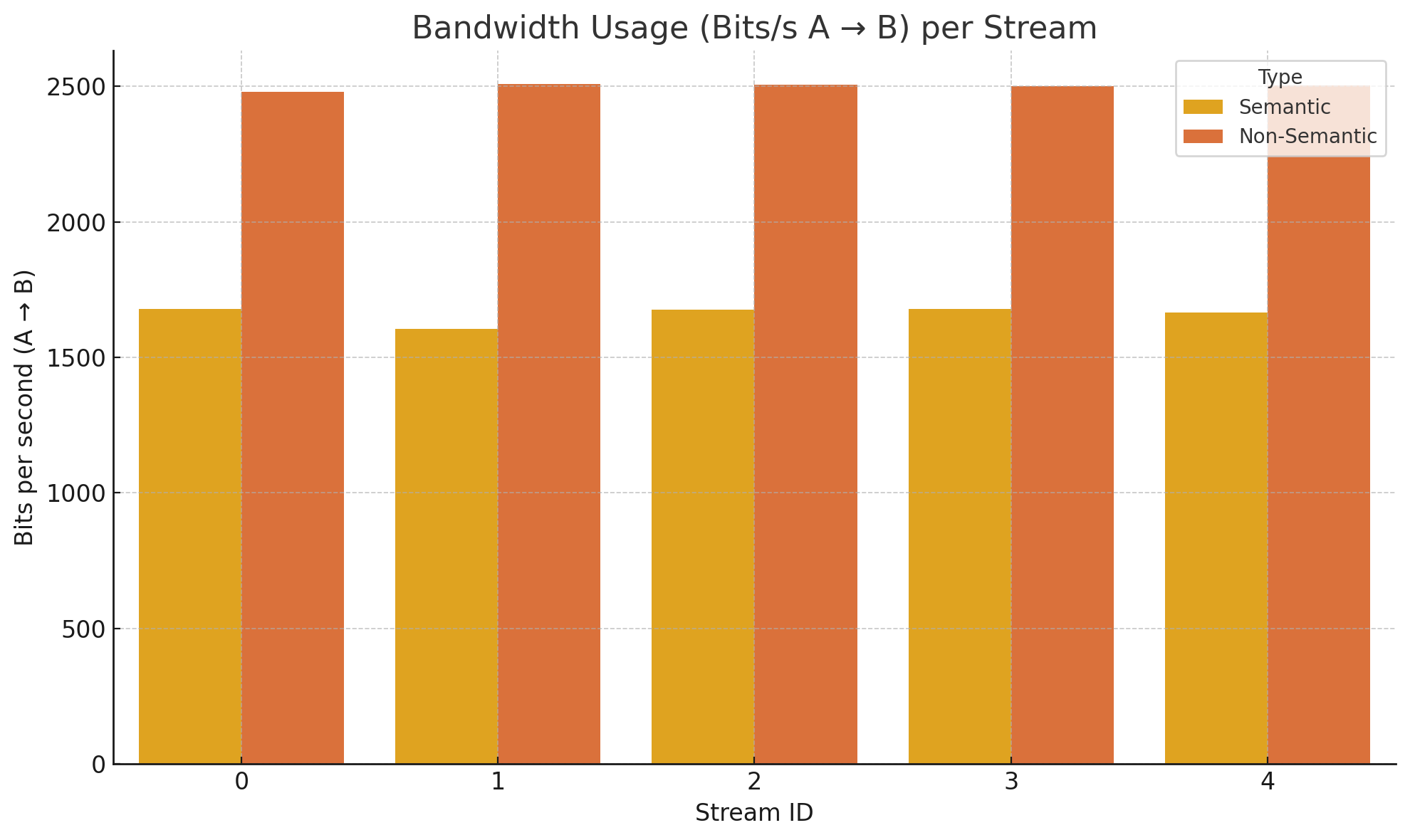}
\caption{Per-Stream Bandwidth Usage in Semantic and Non-Semantic Modes (Vehicle to Server)}
\label{fig:bandwidth-stream}
\end{figure}

Figure~\ref{fig:bandwidth-stream} shows the bits per second transmitted in each stream. In all cases, semantic messages resulted in substantially lower bandwidth usage than their non-semantic counterparts. The semantic streams averaged approximately \textbf{1,661 bits/s}, while non-semantic streams averaged around \textbf{2,500 bits/s}.
This consistent performance across all streams confirms that the bandwidth savings observed are not confined to specific vehicles or conditions but are instead an inherent advantage of semantic communication in our architecture.

\textcolor{black}{ \subsection{Extended Evaluation of Decision Behavior}}

\textcolor{black}{ In addition to bandwidth reduction, we analyze the behavior of the proposed framework from a decision-making perspective.}

\begin{figure}[H]
    \centering
    \includegraphics[width=0.5\linewidth]{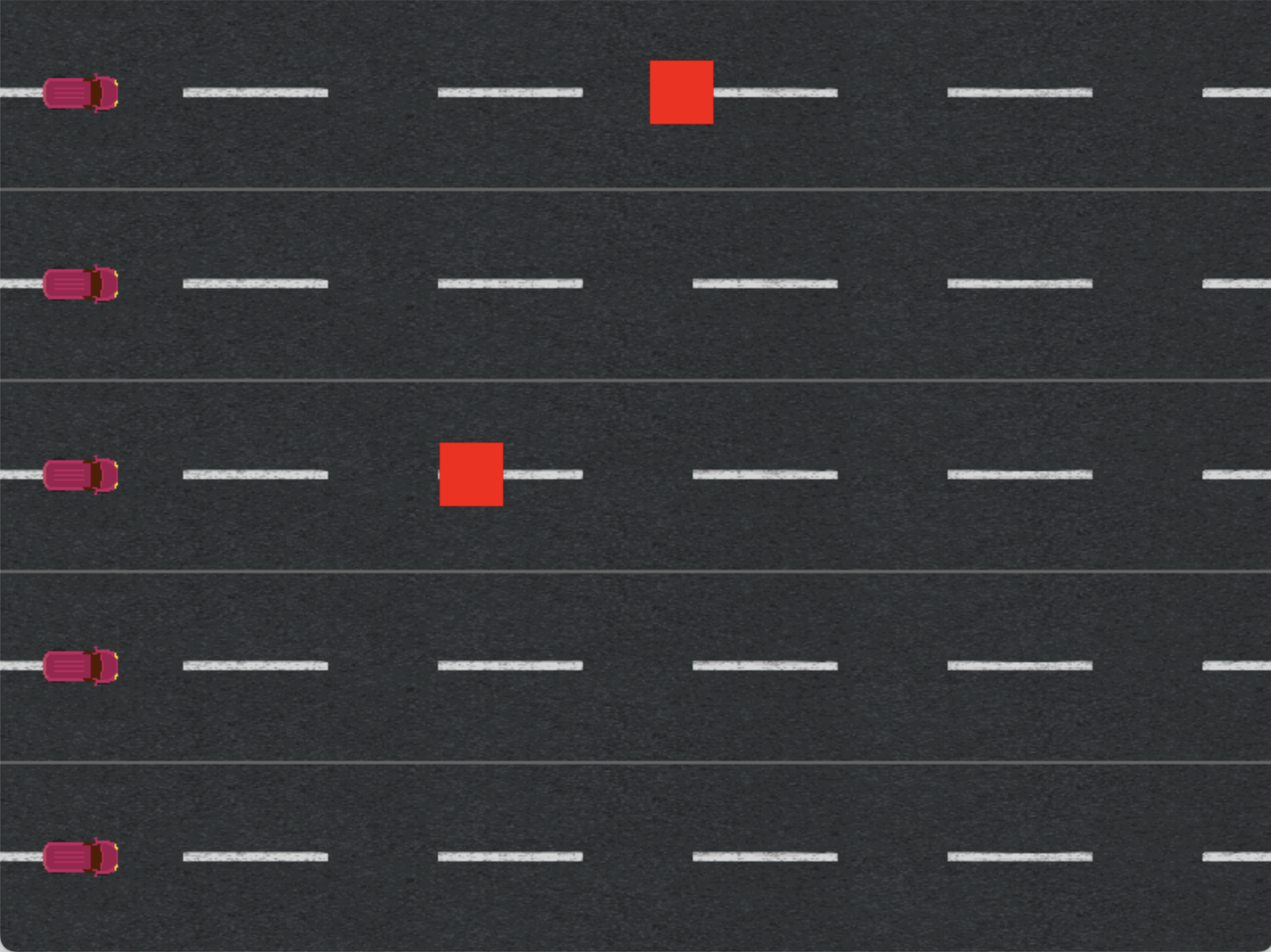}
    \caption{Initial simulation setup with five cars placed across lanes}
    \label{fig:sim_start}
\end{figure}

\begin{lstlisting}[caption={\textcolor{black}{Initial control commands issued by the server to all vehicles under normal driving conditions. This serves as the baseline state before obstacle-triggered adaptation.}},label={list3}, captionpos=b]
Car 1 - MOVE_FORWARD 10
Car 2 - MOVE_FORWARD 10
Car 3 - MOVE_FORWARD 10
Car 4 - MOVE_FORWARD 10
Car 5 - MOVE_FORWARD 10
\end{lstlisting}

\setlength{\parskip}{0pt}

\begin{figure}[H]
    \centering    \includegraphics[width=0.5\linewidth]{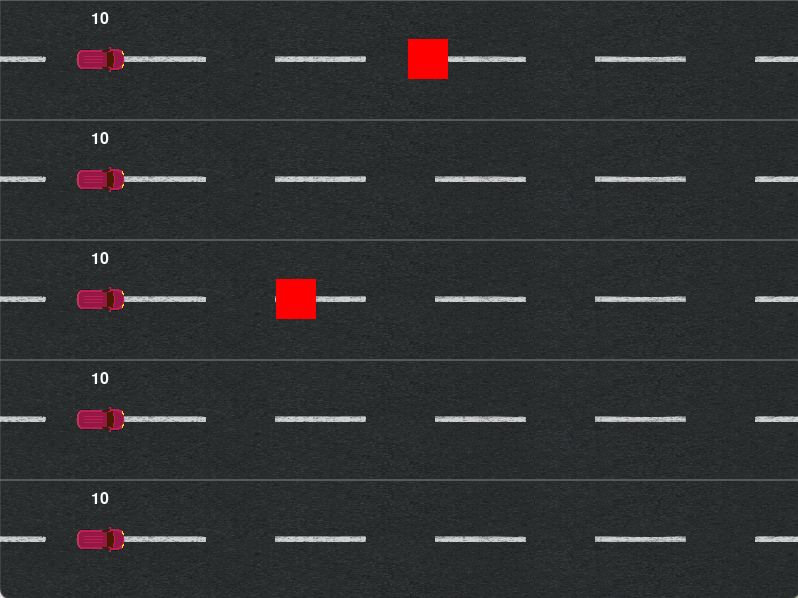}
    \caption{All vehicles proceed forward by 10 units as per the instruction}
    \label{fig:sim_no_obstacle}
\end{figure}

\begin{figure}[H]
    \centering    \includegraphics[width=0.5\linewidth]{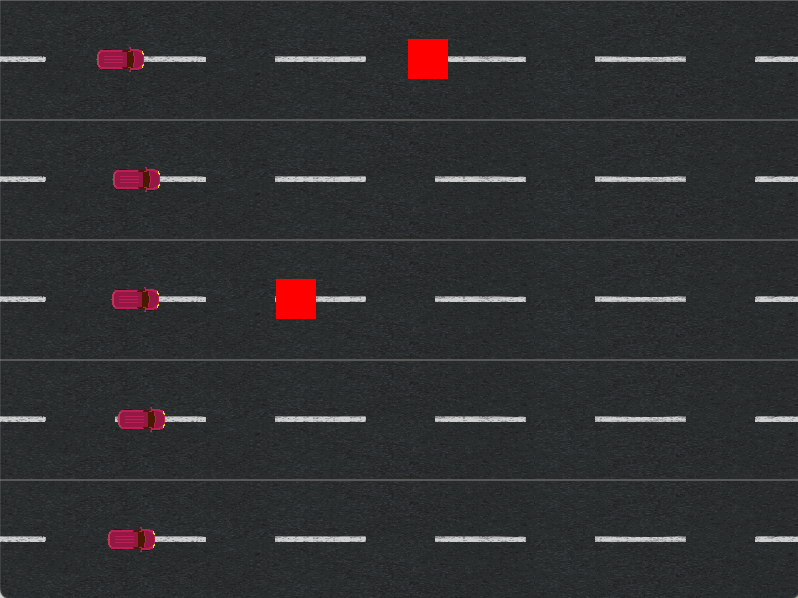}
    \caption{Car 3 approaches an obstacle, while other lanes remain clear}
    \label{fig:sim_obstacle}
\end{figure}

\begin{figure}[H]
    \centering
    \includegraphics[width=0.5\linewidth]{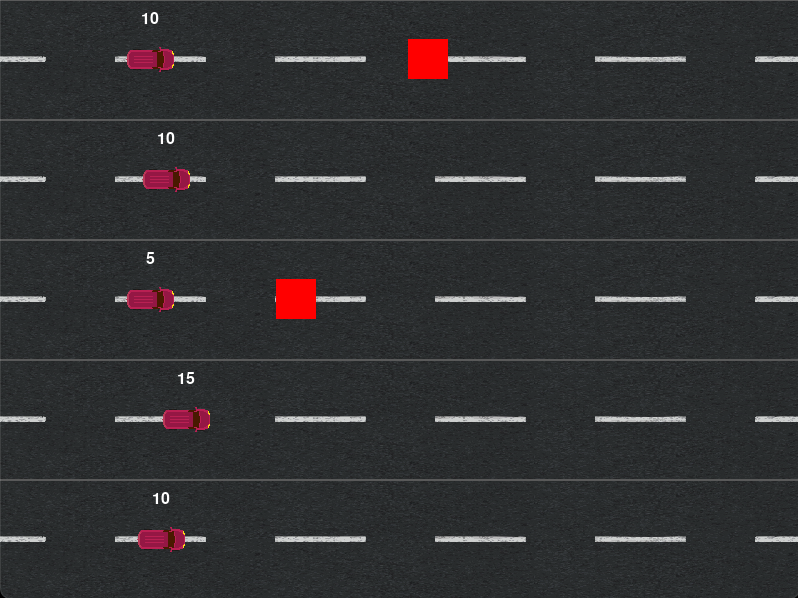}
    \caption{Car 3 slows down slightly while other cars move forward, allowing space adjustment}
    \label{fig:sim_speedup}
\end{figure}

\begin{figure}[H]
    \centering
    \includegraphics[width=0.5\linewidth]{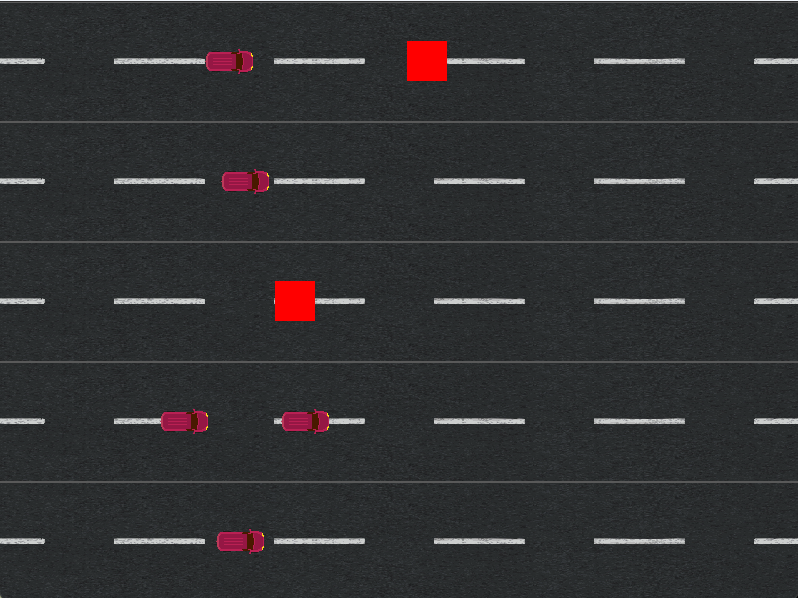}
    \caption{Car 3 now maintains a consistent distance from Car 4, following at 30 units}
    \label{fig:sim_following}
\end{figure}

\textcolor{black}{ \textbf{Scenario-based validation:}
We evaluate the system using a representative lane-change scenario involving five vehicles. As shown in Figures ~\ref{fig:sim_start},~\ref{fig:sim_no_obstacle},~\ref{fig:sim_obstacle}, \ref{fig:sim_speedup} and ~\ref{fig:sim_following}; and Listings~\ref{list3},~\ref{list4},~\ref{list5} and ~\ref{list6} when an obstacle is detected by Car 3, the system generates adaptive control actions by coordinating multiple vehicles. Specifically, Car 3 reduces speed while Car 4 accelerates, creating a safe gap. Once a sufficient distance is established, the system updates its semantic representation by replacing the \textit{SPEED\_UP} relation with a \textit{FOLLOWING} relation at a distance of 30 units. This demonstrates that the system produces context-aware and coordinated decisions based on semantic reasoning.}

\setlength{\parskip}{0pt}

\begin{lstlisting}[caption={\textcolor{black}{Semantic database update generated after obstacle detection. The \texttt{SPEED\_UP} relation is introduced to create a safe gap for coordination.}}, label={list4},captionpos=b]
"relationships": [
  {
    "from_car": "3",
    "to_car": "4",
    "type": "SPEED_UP",
    "properties": {}
  }
]
\end{lstlisting}

\begin{lstlisting}[caption={{\textcolor{black}{Coordinated control commands generated by the server in response to the detected obstacle. The commands adjust vehicle speeds to create safe spacing.}}},label={list5}, captionpos=b]
Car 1 - MOVE_FORWARD 10
Car 2 - MOVE_FORWARD 10
Car 3 - MOVE_FORWARD 5
Car 4 - MOVE_FORWARD 15
Car 5 - MOVE_FORWARD 10
\end{lstlisting}

\begin{lstlisting}[caption={\textcolor{black}{Updated semantic relationships after coordination. The temporary \texttt{SPEED\_UP} relation is replaced by a stable \texttt{FOLLOWING} relation with a safe distance of 30 units.}},label={list6}, captionpos=b]
"relationships": [
  {
    "from_car": "3",
    "to_car": "4",
    "type": "FOLLOWING",
    "properties": {
      "distance": 30
    }
  }
]
"delete_relationships": [
  {
    "from_car": "3",
    "to_car": "4",
    "type": "SPEED_UP",
    "properties": {}
  }
]
\end{lstlisting}

\textcolor{black}{ \textbf{Illustrative safety-aware decision behavior:}
In the representative scenario, the generated actions respect predefined safety constraints embedded in the decision logic, including collision avoidance, safe inter-vehicle distance, and lane-change feasibility. In the evaluated scenarios, no unsafe actions (e.g., conflicting lane changes or unsafe proximity) are observed in the illustrated scenario, indicating that semantic reasoning aligns with safe driving behavior.}
\vspace{0.23cm}

\textcolor{black}{ \textbf{Decision consistency:}
We observe that under similar traffic conditions, the system produces consistent control decisions (e.g., repeated MOVE\_FORWARD or adaptive speed adjustments), suggesting stable behavior of the LLM-based reasoning process within the simulation environment.}

\vspace{0.23cm}
\textcolor{black}{ \textbf{Semantic efficiency:} The results complement the bandwidth analysis by showing that semantic messages not only reduce data transmission but also preserve task-relevant information required for decision-making, enabling efficient and interpretable coordination among vehicles.}
\textcolor{black}{
To illustrate the difference between raw and semantic communication, Listings 1 and 2 present an example of a LIDAR-based JSON payload and its corresponding semantic representation. The semantic message captures only task-relevant information (e.g., obstacle presence and relative position), reducing redundancy while preserving actionable meaning. In the illustrated example of Listings 1 and 2, the semantic representation is approximately 2–3× shorter than the corresponding raw JSON payload; this comparison is intended only as an example of message compactness, while the main quantitative result is the overall 33.54\% bandwidth reduction reported in the experiments.
}

\textcolor{black}{ These results provide an initial indication of decision reliability and system-level behavior. However, we note that formal evaluation using metrics such as collision rate, trajectory stability, and real-time latency is beyond the scope of this prototype and is left for future work.}

\section{Conclusion}
\label{sec:Conclusion}

This paper introduced an architecture that combines semantic communication with LLMs to improve decision-making and communication efficiency in V2X systems. The proposed system addresses key limitations in traditional vehicular communication frameworks, such as bandwidth inefficiency and context-insensitive reasoning, by abstracting raw sensor data into semantically meaningful messages and leveraging LLM for real-time interpretation and control. Through the integration of a graph-based memory using Neo4j and the semantic inference capabilities of GPT-4, the system demonstrated substantial improvements in bandwidth usage.
Experimental results showed that semantic messaging reduced total transmission volume by about 33.54\% compared with a non-semantic baseline, \textcolor{black}{ while demonstrating coherent and context-aware decision behavior in the evaluated scenarios.} Structuring environment information in a shared semantic space also improved coordination between vehicles and infrastructure, making interactions more scalable and interpretable.

Future work will focus on validating the framework in more complex settings (e.g., high-fidelity simulators like CARLA and, eventually, physical robotic platforms) to study scalability, hardware integration, and safety constraints.

\section*{Acknowledgment}
This work is supported in part by the EU MSCA Project “COALESCE”
under Grant Number 101130739, by the US-Ireland R\&D Partnership Programme RI-SFI-23/US/3924, and by Science Foundation Ireland Grant 13/RC/2077\_P2.

\vfill

\end{document}